# On the structure of quantum intermediate state in type I superconductors

O. P. Ledenyov

*National Scientific Centre Kharkov Institute of Physics and Technology,
Academicheskaya 1, Kharkov 61108, Ukraine.*

The calculation of spatial structure of a quantum intermediate state in high pure *type I* superconductors in external magnetic field at low temperatures is completed. The theoretical model of thermodynamics of intermediate state in *type I* superconductors was proposed by Andreev. It is shown, that in the quantum case, the period of structure of quantum intermediate state appears to be significantly shorter and it has different dependences on the magnetic field $H$ and temperature $T$ in comparison with the classical intermediate Landau state. The decrease of thickness of normal layers results in the increase of characteristic distance between the quantum Andreev levels of electronic excitations, and leads to the transition to the quantum intermediate state from the classical intermediate Landau state, which is realized at the higher temperatures $T \sim 1$ K. The comparison of the computed data with the experimental results in the researched case of high pure *Gallium* single crystal is made.

## Introduction

The theory of intermediate state (*IS*) in *type I* superconductors was created by Landau [1] in frames of the classical electrodynamics by Gorter and Casimir (see, for example, [2]). In agreement with the Landau theory, there is the intermediate state (*IS*) in *type I* superconductors at action of external magnetic field at low temperatures, when the following expression is true: $(1-n)H_c < H < H_c$. The *IS* structure has a periodic layered structure with the interchangeable normal metal and superconductor layers ($n$ is the demagnetization factor of a superconducting sample, which depends on the superconducting sample's geometric form and its orientation in magnetic field). In [1], the spatial period of *IS* structure in superconducting plate in transverse magnetic field was found, and the *IS* structure dependence on the samples geometric dimensions and the surface energy, appearing on the boundary, created by the normal and superconducting phases, was researched. Subsequently, Landau developed the theory on the *IS* structure with the layers branching in *type I* superconductors [3]. In [4], Lifshitz and Sharvin analyzed the theories [1, 3], deriving the convenient formulas for a comparison with the experimental results. The Landau theory and subsequent experimental results, occurred to be in good agreement with the phenomenological theory of superconductivity by Ginzburg - Landau [5] and the microscopic theory by Bardeen, Cooper, Schrieffer (*BCS*) [6], and are well described in the textbooks and monographs (see, for example, [2, 7, 8]). Among more recent theoretical works, let draw attention to the detailed research on the *IS* structure properties, including the calculation on the goffered twisted *IS* structures, observed in superconductors at specific conditions in [9]. It is interesting to note that, in this case, the energy of interaction between the currents, flowing through the boundaries between the *S-N-S* layers, plays a main role in the creation of *IS* structures. The Landau theory and following research works, which presented the foundation for the creation of *IS* structure theory, are based on the classical case representation, which is in a good agreement with the experimental findings in the case of *type I* superconductors in external magnetic field at the temperatures $T \geq 1$ K.

In [10, 11], Andreev attracted attention to an importance of the quantum effects in the *IS* structure creation in superconductors. In [10], the specific mechanism to explain the electronic excitations (*EE*) reflections in normal metal layers on the boundaries, created by the normal and superconducting phases, was introduced. These *EE* reflections are accompanied by the *EE* transition from the electron spectrum to the hole spectrum (and vice versa) with the change of their movement direction on the opposite direction, and were experimentally confirmed to exist in the *S-N-S* systems. In [11], it was shown that, at low temperatures, when the mean free path by the *EE* is bigger than the thickness of normal layer $l >> d_n$, the *EE* are captured in the normal metal layers as a result of the *EE* reflections on the *N-S* boundaries, leading to the spatial quantization of the *EE* spectrum.

The distance between the Andreev levels is

$$\Delta\varepsilon = \frac{h v_F \cos\theta}{d_n}, \qquad (1)$$

where $h$ is the Plank constant, $v_F$ is the velocity of the *EE* on the Fermi surface. As it is visible, the energy depends on the thickness of the normal metal layer $d_n$ and the angle $\theta$ between the direction of movement by the *EE* and the perpendicular to the normal metal - superconductor (*N-S*) boundary. In agreement with [11], in the superconductors, in which the mean free path by the *EE* is significantly bigger than the thickness of normal layer $l \gg d_n$, at the low temperature, which is smaller than some characteristic temperature

$$T^* = \frac{h v_F}{k_B d_n}, \qquad (2)$$

the Andreev quantization has a considerable influence on the main thermodynamic characteristics of *IS* structure(the free energy, thermal capacity, magnetic momentum) and makes them depending on the thickness of the normal metal layer $d_n$ ($k_B$ is the Boltzmann constant). Going from the thickness of normal layer $d_n$, which is characteristic for the classical *IS* structure, the magnitude of transition temperature $T^* \approx 0,1$ K was found in [11]. At the same time, the problem about a possible influence by the *EE* spectrum quantization on the *IS* structure was not researched in [11].

Let note that, after the publication of research results by Andreev [11], Zavaritsky [12] discovered that, in the *IS* in the superconducting *Tin* at the temperature $T<0.18$ K, the *QIS* structure properties in a dependence of the thermal capacity on the temperature are observed in agreement with the prediction by Andreev [11]. At later date, the quantitative research on the *IS* structure in a superconducting high pure *Gallium* single crystal in external magnetic field at the low temperature $T \approx 0.3$ K was conducted, and the dependence of the thickness of normal metal layers on the magnetic field was found by the ultrasound attenuation measurement method by Ledenyov, Fursa [13]. In [13], going from the theory by Andreev [14] for the monotone part of ultrasound amplitude attenuation dependence and the theory by Ledenyov [15] for the oscillating part of ultrasound amplitude attenuation dependence, the obtained experimental results were analyzed, leading to the conclusion that there is a new *IS* structure in *type I* superconductor in external magnetic field at low temperatures, which has the small thickness of normal metal layers in distinction from the well known classical Landau *IS* structure in *type I* superconductors in external magnetic field at low temperatures. It was difficult to come up with the idea that the new *IS* structure in a superconducting high pure *Gallium* single crystal in external magnetic field at low temperature represents the *QIS* structure, because the nature of the new *IS* structure was not fully understood on that time.

In this research, the calculations with the appropriate formulas derivation to describe the quantum intermediate state (*QIS*) structure in high pure *type I* superconductors in external magnetic field at low temperatures, using the thermodynamic theory [11], are completed. It is clarified that, in the quantum case, the period of *QIS* structure appears to be much shorter, having the different dependences on the magnetic field *H* and the temperature *T* in comparison with the classical Landau intermediate state structure. It is shown that the characteristic temperature of transition to the *QIS* in high pure *type I* superconductors in external magnetic field is higher, comparing to the temperature of transition to the classical Landau intermediate state (*IS*) in *type I* superconductors in external magnetic field [11], because of the small thickness of normal metal layers in the *S-N-S* layered structure in the quantum intermediate state in high pure *type I* superconductors in external magnetic field at low temperatures. The obtained theoretical calculations data are in good agreement with the ultrasound propagation and attenuation experimental measurements results in the high pure *Gallium* single crystal in external magnetic field at low temperatures [13].

The research on the electron properties as well as the superconducting properties of these systems is of certain interest, because of the synthesis of nano-structurized composite materials, in which the quantum effects may appear at higher temperatures, having considerable influence on their thermodynamic and structural properties.

## Quantum intermediate state structure in type I superconductors

The thermodynamic characteristics of intermediate state in superconductors depend on the external magnetic field, therefore the Gibbs free energy is a convenient thermodynamic potential for their accurate characterization [7]. Let emphasis that the potential *G* takes to the account the work, done by the source of external magnetic field, hence the signs of corresponding contributions are defined by this circumstance. Let write, at the temperature $T<T_C$, the free energy of a superconducting sample with volume *V*, which is in the normal metal (3) and superconducting (4) states in the external magnetic field $H<H_C(T)$

$$G_{N,H} = VF_{N,0} - VH^2/8\pi - V_{ext}H^2/8\pi, \qquad (3)$$

$$G_{S,H} = VF_{S,0} - V_{ext}H^2/8\pi \qquad (4)$$

where *V* is the internal volume of a superconducting sample, $V_{ext}$ is the external volume of system with magnetic field, $F_{N,0} - F_{S,0} = H_C^2/8\pi$, $H_C$ is the critical magnetic field at the temperature of *T*. Let write the potential of unit of volume of superconductor, which transited to the intermediate state (*IS*) at the applied external magnetic field *H*, assuming that some part of it's volumetric fraction $\eta$ is in the normal metal state. Let count the potential, starting with its value in the superconducting state

$$G_{I,H} = \eta F_{N,0}(d_n) - \eta H_N^2/8\pi - \eta F_{S,0} - G_{surf}(d_n), \qquad (5)$$

where $H_N$ is the magnetic field in the normal metal layers of a superconducting sample, which may be different from the magnitude of critical magnetic field $H_C$. The volumetric inputs by the external regions, situated far away from a sample, don't depend on its state, hence they are mutually reduced. At the same time, the inputs by the internal regions with inhomogeneous magnetic field at internal boundaries, dividing the superconducting and normal metal layers, as well as the inputs by the regions with inhomogeneous magnetic field near to the surface of a sample, connected with the discrete nature of the *IS* structure, have to be taken to the consideration. These inputs can be written as the term $G_{surf}$, which is equal to the work by external source to create the boundary inhomogeneities of magnetic field in eq. (5). In frames of the Landau theory, during the finding of an extreme of the potential $G_{I,H}$, the term $G_{surf.}$ depends on the period of *IS* structure only. Let note that the concentration of the normal phase $\eta$ doesn't depend on the period of *IS* structure in a sample. Let move from the description of the free energy *G* as a function of the external source to the description of the free energy *G* as a function of the sample's state as in the Landau theory [1]. Thus, all the terms will change their signs, and the term $G_{surf.}$ will have a positive value, and the period of the *IS* structure will be defined by the minimum of potential. In the quantum approach, in distinction to the classical Landau theory, it is necessary to consider the first term in expression (5), which is connected with the volumetric electron energy by the *EE*, which is a function of the thickness of normal metal layer in this case [11].

In the Landau theory, the two energy inputs, contributing to the density of free energy of a sample, are taken to the account (calculated on the unit of cross-section, which is transverse to the external magnetic field, and counted, starting from the free energy of a superconductor). The two energy inputs depend on the period *d* of *IS* structure and the thickness of the normal metal layers $d_n$ as $G_{surf}(d_n) = F_1(d_n)+F_2(d_n)$. The two energy inputs $F_1$ and $F_2$ are present in the quantum theory.

The first energy input $F_1$ by the superconducting and normal phases separation boundary is

$$F_1 = 2L\delta\, H_C^2 \big/ 8\pi d\, , \qquad (6)$$

where *d* is the period of *QIS* structure, $d = d_n + d_S$, $d_n = \eta\, d$, *L* is the thickness of a plate of superconductor, $\delta$ is the parameter of surface energy at *N-S* boundary, $\delta = \xi - \lambda$, $\xi$ is the correlation length, $\lambda$ is the penetration depth by magnetic field into a superconductor.

The second energy input $F_2$ is connected with the energy of inhomogeneity of magnetic field near to the surface of a sample. The exact calculation of the value of second input is based on the approach, which considers the geometry of the *S-N-S* layers near to the surface of a superconductor in the research by Lifshits and Sharvin [4], and it is

$$F_2 = \varphi(\eta)d\, H_C^2 \big/ 8\pi\, , \qquad (7)$$

where $\varphi(\eta)$ is the tabulated function.

Using a different approach in [7], the second energy input $F_2$ can be approximately calculated as a function, expressed by the concentration of normal phase $\eta$ and by the period *d* of *IS* structure. However, the accuracy of the result, obtained with the use of expression in [7], must be revised due to its exceeding value in comparison with the actual magnitude $\varphi(\eta)$ (in three times bigger approximately). This problem is solved in the below provided calculations.

Going from the fact that the magnitudes of magnetic field in the outside and inside domains of a sample in the intermediate state are different, let assume that there is a transition region with the effective length $\lambda_{eff}$ at which the tuning of the external magnetic field $H_e$ and internal magnetic field $H_N$ to some equilibrium magnitude of magnetic field takes place. Obtaining an average of the magnitude of magnetic field on the distance, which is longer than the period of the *IS* structure, let follow the general rule of change of magnitudes difference of magnetic fields in the outside and inside regions of a superconductor near to its boundary as *exp(-kx)*, where *x* is the distance from the surface of a sample, and $k=2\pi/\lambda_{eff}$. It is convenient to express the spatial parameter $\lambda_{eff}$ as in [7]

$$1/\lambda_{eff} = 1/d_n + 1/d_S = d/d_n d_S\, . \qquad (8)$$

One can see that the magnetic field changes inside the sample as well as outside the sample on the distance $x=1/k=\eta(1-\eta)d/2\pi$. There is no sharp change of magnitude of magnetic field in the normal metal layer on the surface of a sample. There is a fluent change of the intensity of magnetic field, occurring in the outside and inside regions of a sample due to an appearance of currents on the *N-S* boundary in the superconducting layers in a sample. Inside the sample, the domain of magnetic field change spreads on the distance around *1/k*, and it is connected with the decrease of thickness of superconducting layers at approach to the surface of a sample; outside the sample, it is connected with the presence of currents on the boundaries of superconducting layers in this part of a sample. The excessive density of magnetic energy in the region of inhomogeneity of magnetic field can be assessed in agreement with [7], as a difference between the average density of energy in normal metal layers $\eta H_N^2/8\pi$ and the density of average magnetic field $(\eta H_N)^2/8\pi$, hence it is equal to $\eta^2(1-\eta)^2 H_N^2/8\pi$. The value of additional input energy $F_2$, originated by the *S-N-S* layered structure appearance on the surface of a sample, is

$$F_2 = 4d\left[\eta^2\left(1-\eta^2\right)\big/2\pi\right]H_N^2\big/8\pi\, , \qquad (9)$$

where $H_N$ is the magnetic field in a layer with normal phase, which is close to the critical magnetic field in general case. The approximate analytic expression for the function $\varphi(\eta)$ in the Landau theory is

$$\varphi(\eta) \approx \eta^2(1-\eta^2)\big/\pi\, . \qquad (10)$$

Neglecting the difference between the magnetic fields $H_C$ and $H_N$, and minimizing the sum of the inputs $F_1$ and $F_2$ by the period *d*, the well known expression for the period in the Landau – Lifshits - Sharvin theory can be obtained

$$d = \left[L\delta\big/\varphi(\eta)\right]^{1/2}\, . \qquad (11)$$

In agreement with the completed calculations, the Lifshits - Sharvin function $\varphi(\eta)$ can be approximated with great precision, than in the eq. (10), and presented in the form of the convenient expression

$$\varphi(\eta) \approx \frac{\eta^2(1-\eta^2)}{\pi}\ln\left[1+\frac{\eta+\pi/4}{\eta\left(1+\eta/e+\eta^2/e\right)}\right], \quad (12)$$

where $e \approx 2{,}7183$ is the base of natural logarithm. The expression (10) will be used with the purpose to simplify the formulas for the dependence $\varphi(\eta)$.

In addition to the sum of the energy inputs $F_1$ and $F_2$, the volumetric input $\eta F_{N,0}$, which depends on the period $d$ and the thickness of the normal metal layer $d_n$ in the QIS structure in the quantum case in distinction from the classical case, has to be taken to the account during the calculation of the period $d$ of QIS structure. The energy $F_A$ as a function of the period $d$ of QIS structure can be written as

$$F_A = \alpha\,\beta\,C_{el}\eta^2 d\,L\,T^2/v_F^2, \quad (13)$$

where $\alpha$ is the constant, which is close to $1$, as calculated in [10], $\beta$ is the coefficient, which takes to the account the deviation of the Fermi surface from the isotropic surface for which $\beta=1$, and the fraction of phase volume of electronic excitations, covered by the quantization, $C_{el} = \gamma T$ is the thermal capacity of normal metal without consideration of Andreev quantization. Minimizing the total density of free energy in the IS structure with the consideration of $F_A$, we will derive the expression for the period $d$

$$d = \frac{(2L\delta)^{1/2}}{\left[2\eta^2(1-\eta)^2/\pi+(\alpha\beta\gamma\eta^2 LT^3/v_F^2)/(H_C^2/8\pi)\right]^{1/2}} \quad (14)$$

The thickness of normal metal layer is simply connected with the period of QIS structure $d_n = \eta d$, and it can be written as

$$d_n = \left(2L\delta\big/\left[(2(1-\eta)^2/\pi)+8\pi\alpha\beta\gamma T^3 L/v_F^2 H_C^2\right]\right)^{1/2}. \quad (15)$$

One can see that the derived formulas for the QIS period $d$ (14) and the thickness of the layer of normal phase $d_n$ (15) in the case of QIS structure are different from the $d$ and $d_n$ expressions in the case of Landau IS structure (the formula (15) was also used in [16]). The presence of additional term in denominators in these expressions results in the significant changes in the shapes of dependences of QIS structure period on the external magnetic field and temperature $d(H, T)$, and the thickness of the layer of normal phase on the external magnetic field and temperature $d_n(H, T)$. The system of S-N-S layers transforms into the periodic QIS structure with the significantly shorter period $d$ and much thinner thickness of normal metal layers $d_n$ with the purpose of minimization of the Gibbs free energy. The surface energy on the normal metal - superconductor phases separation boundaries counteracts to the decrease of the QIS structure period $d$, influencing in accordance with the Landau theory. Let note that, in the considered case, the value of the period $d$ is small and the energy of magnetic field inhomogeneity adds a small input to the free energy of a sample, that is why the energy, connected with the Andreev quantization, has a main impact on the system. The QIS structure period has the weak dependences on the external magnetic field $H$ and the concentration of normal phase $\eta=(H-(1-n)H_C)/nH_C$.

In *Fig. 1*, the corresponding graphs with the computed dependences of the period of structure $d$ (1) and the thickness of normal metal layer $d_n$ (2) on the concentration of the normal phase $\eta$ in a quantum intermediate state in the high pure *Gallium* single crystal at the temperature of *0.35 K*, are presented. The characteristic magnitudes of parameters, measured in the case of *Gallium* single crystal in [13], have been used during the computing modeling.

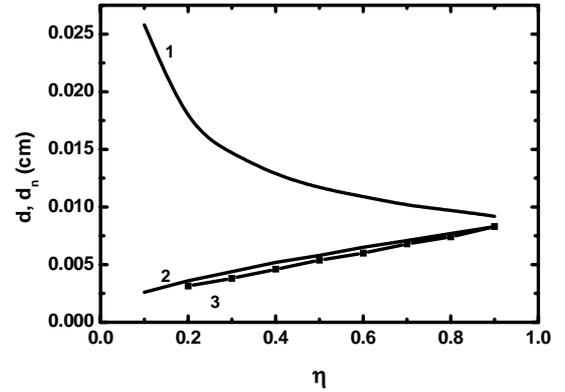

*Fig. 1. Computed dependences of period of structure d (1) and thickness of normal metal layer $d_n$ (2) on concentration of normal phase $\eta$ in quantum intermediate state. The curve (3) shows the experimental results, obtained during the research on ultrasound attenuation in high pure Gallium single crystal at T=0.35 K [13].*

As it is visible in *Fig. 1*, the period of QIS structure does not increase, when the concentration of normal phase $\eta$ approaches to $1$. At the same time, in agreement with (11), in the classical case, at $\eta \to 0$ and at $\eta \to 1$, the period of IS structure must increase sharply. The thickness of normal metal layer $d_n$ must sharply increase at $\eta \to 1$ in the case of the IS structure, while the dependence $d_n(\eta)$ is close to linear in the QIS structure. As it is shown in *Fig. 1*, the experimental data, obtained during the research on the ultrasound propagation and attenuation in high pure *Gallium* single crystal in magnetic field at low temperatures, are in good agreement with the theoretical modeling results, using the formulas, based on the proposed theory.

The dependence of the thickness of normal metal layer on the temperature $d_n(T)$ in the case of QIS differs from the similar dependence in the classical IS in *type I* superconductors in external magnetic field at low temperature as it follows from eq. (15).

## Conclusion

In this research, the theoretical calculations on the spatial structure of quantum intermediate state, appearing in high pure *type I* superconductors in external magnetic field at low temperatures, in which the electron excitations have the long mean free path $l$, which is bigger than the thickness of the normal metal layer $l > d_n$, are completed. It was found that, at low temperatures, the high pure *type I* superconductors have transition to the quantum intermediate state, which is characterized by the thinner normal metal layers in comparison with the thickness of normal metal layers in the classic intermediate Landau state. The contribution by the normal electronic excitations to the free energy is significantly smaller in the case of thinner normal metal layers in quantum intermediate state in high pure *type I* superconductors in external magnetic field at low temperatures. It is shown, that the period of structure of quantum intermediate state appears to be significantly shorter in comparison with the period of structure of classical intermediate Landau state. The period of structure of quantum intermediate state has different dependences on the magnetic field $H$ and temperature $T$ in comparison with the period of structure of classical intermediate Landau state. The theoretical calculations data, obtained in this work, perfectly coincide with the experimental results for the high pure *Gallium* single crystal in external magnetic field at low temperatures [13]. In author's opinion, the *type I* superconductors of high purity with the long mean free paths by electronic excitations $l > d_n$ in external magnetic field at low temperatures represent a new class of quantum objects with the unique physical properties, which have to be researched in details.

Author thanks V.O. Ledenyov and D.O. Ledenyov for the discussion on the results of research.

This article was published in Russian in the Problems of Atomic Science and Technology (*VANT*) [17].

———————

1. L.D. Landau, Zh. Eksp. Teor. Fiz. **7**, 371 (1937) (in Russian).
2. D. Shoenberg, *Superconductivity*, IIL, Moscow, 288 (1955) (in Russian).
3. L.D. Landau, Zh. Eksp. Teor. Fiz. **13**, 377 (1943) (in Russian).
4. E.M. Lifshitz, Yu.V. Sharvin, Dokl. Akad. Nauk USSR **79**, 783 (1951) (in Russian).
5. V.L. Ginzburg, L.D. Landau, Zh. Eksp. Teor. Fiz. **20**, 1064 (1950) (in Russian).
6. J. Bardeen, L.N. Cooper, J.R. Schrieffer, Phys. Rev. **108**, 1175 (1957).
7. M. Tinkham, *Introduction to Superconductivity*, Chapter 3, McGraw-Hill Book Comp. (1975).
8. J.D. Livingston, W. Desorbo, *The Intermediate State in the Type I Superconductors*, in *Superconductivity*, v. 2, ed. by R.D. Parks, Marcel Dekker Inc., N.Y., 1235 (1969).
9. A.T. Dorsey, R.E. Goldstein, Phys. Rev. B **57**, 3058 (1998).
10. A.F. Andreev, Zh. Eksp. Teor. Fiz. **46**, 1823 (1964) (in Russian).
11. A.F. Andreev, Zh. Eksp. Teor. Fiz. **49**, 655 (1965) (in Russian).
12. N.V. Zavaritsky, Pisma Zh. Eksp. Teor. Fiz. **2**, 168 (1965) (in Russian).
13. O.P. Ledenyov, V.P. Fursa, Fiz. Niz. Temp. **11**, n. 1, 57 (1985) (in Russian).
14. A.F. Andreev, Zh. Eksp. Teor. Fiz. **53**, 680 (1967) (in Russian).
15. O.P. Ledenyov, Pisma Zh. Eksp. Teor. Fiz. **30**, 185 (1979) (in Russian).
16. O.P. Ledenyov, V.O. Ledenyov, D.O. Ledenyov. *Quantum Effects in Type I Superconductors in Magnetic Field*, Proceedings of International Conference "Physics of Condensed Matter at Low Temperatures", NSC KIPT, Kharkov, Ukraine, 82, (2006).
17. O.P Ledenyov, Problems of Atomic Science and Technology (*VANT*), no. 1(17), 48 (2008) (in Russian);
http://vant.kipt.kharkov.ua/ARTICLE/VANT_2008_1/article_2008_1_48.pdf .